\documentclass[10pt,draft]{dis03}
\usepackage{epsf,amsmath}

\def\bea{\begin{eqnarray}}
\def\eea{\end{eqnarray}}

\newcommand{\as}{\alpha_{\mathrm{s}}}
\newcommand{\ee}{e^+e^-}
\newcommand{\cF}{{\cal{F}}}    %   the multiple emission correction factor
\newcommand{\muf}{\mu_\textsc{f}}

\newcommand{\CF}{C_F}
\newcommand{\TR}{T_R}
\newcommand{\CA}{C_A}

\newcommand{\nf}{n_{\!f}}

\def\cO#1{{\cal{O}}\left(#1\right)}
\begin{document}

\title{AUTOMATED RESUMMATION OF \\JET OBSERVABLES IN QCD}

\author{GIULIA ZANDERIGHI \\
IPPP and University of Durham\\ 
E-mail: giulia.zanderighi@dur.ac.uk }
\maketitle

\begin{abstract}
\noindent 
We present a master formula, with applicability conditions, which
allows us to automate the resummation of infrared and collinear
logarithms appearing in distributions of jet observables in QCD at
next-to-leading logarithmic accuracy.
\end{abstract}
\section{Jet observables: fixed order and resummation}
Jet observables, event shapes and jet rates, revealed themselves as
one of the richest laboratories to explore QCD.
Being infrared and collinear safe (IRC), they can be predicted with
perturbative (PT) techniques, but their high sensitivity to low energy
emissions allows us to investigate the fairly unknown non-perturbative
regime.
Most discriminatory studies make use of distributions. In integrated
distributions $\Sigma(v)$ one requires that the value of the
observable $V(\{\tilde p\}, k_1,\dots,k_n)$ -- a function of all
secondary final state momenta $k_i$ and of the Born momenta after
recoil from the emissions $\{{\tilde p}\}$ -- be less than a fixed
value $v$
\begin{equation}
\label{eq:sigma}
  \Sigma(v) = \int dV \frac{1}{\sigma}\frac{d\sigma}{dV} \, 
\theta(v-V(\{\tilde p\}, k_1,\dots,k_n))\>. 
\end{equation}
The inclusive phase space region, where $v = \cO{1}$, is dominated by
events with hard jets, which move the value of the observable far away
from its Born value, $v=0$.  These events can be described with fixed
order PT expansions, however they are quite rare, every additional jet
being suppressed by an additional factor of $\as$.
More common events are characterised by a large number of
soft-collinear emissions which modify only slightly the value of the
observable from its Born value, so that $v \ll 1$. Here fixed order
predictions fail since every power of $\as$ is accompanied by up to
two large logarithms $L=\ln(1/v)$ of the value of the observable.
A reorganization of the PT expansion is then needed in order to resum
all leading (LL, $\exp\{\as^n L^{n+1}\}$) and next-to-leading (NLL,
$\exp\{\as^n L^{n}\}$) logarithmic terms.

In the past few years the analytical resummation for a variety of
observables have been presented in $\ee$-collision\cite{CTTW,CMW} and
DIS\cite{NG}.  The matching of resummed predictions with fixed order
results allowed tests of QCD, measures of the coupling constant and
studies of non-perturbative corrections\cite{EvShapesExp}.
However, the need for a separate analytical calculation for every
observable has limited the experimental use of resummed predictions.
When dealing with multi-jet observables, analytical calculations
become quite unfeasible, involving many integral transforms in order to
write the distribution in a factorized form\cite{eeKout}. Also in some
cases it turns out not to be possible to resum the observable
analytically\cite{BSZ}.

We present then here a general approach to resummation based on a
preliminary automated analysis of the observable, to establish its
relevant properties with respect to soft-collinear emissions; in a
subsequent step this information is used as an input of a general
master formula.

\section{Automated resummation}
\subsection{Analysis and applicability conditions}
We start by considering a Born event consisting of $n$ hard partons
(`legs'), $n_i$ of which are incoming, with momenta $p_1\ldots p_n$.
We resum $(n\!+\!1)$-jet observables in the $n$-jet limit.  The
(positive defined) observable should then
\begin{enumerate}
\item vanish smoothly as a single extra ($n$+$1$) parton
  of momentum $k$ is made soft and collinear to a leg $\ell$, with the
  functional dependence 
  \begin{equation}
    \label{eq:simple}
    V(\{{\tilde p}\}, k)=
    d_{\ell}\left(\frac{k_t}{Q}\right)^{a_\ell}e^{-b_\ell\eta}
    g_\ell(\phi)\>. 
  \end{equation}
  Here $Q$ is a hard scale of the process and the secondary emission
  $k$ is defined in terms of its transverse momentum $k_t$ and
  rapidity $\eta$ with respect to leg $\ell$, and where relevant, by
  an azimuthal angle $\phi$ relative to a Born event plane. By
  requiring the functional form \eqref{eq:simple} (in practise, almost
  always valid), the problem of analyzing the observable reduces in
  part to identifying, for each leg $\ell$, the coefficients $a_\ell$,
  $b_\ell$, $d_\ell$ and the function $g_\ell(\phi)$.  IRC safety
  demands $a_\ell > \max\{0, -b_\ell\}$.
\item be \emph{recursively} IRC safe, i.\ e.\ given an ensemble of
  arbitrarily soft and collinear emissions, the addition of a
  relatively much softer or more collinear emission should not
  significantly alter the value of the observable, condition required
  for exponentiation of the leading logarithms.  Observables like
  jet-rates in the Jade algorithm are then excluded, and many other
  examples of non-exponentiating IRC safe observables
  exist\cite{autolet+autopaper}.
\item be continuously global\cite{NG} --- this means that for a
  single soft emission, the observable's parametric dependence on the
  emission's transverse momentum (with respect to the nearest leg)
  should be independent of the emission direction. In practice this is
  perhaps the most restrictive of the conditions.  It implies $a_1 =
  a_2 = \ldots = a_n \equiv a$.
\end{enumerate}
To enable our computer program CAESAR 
(Computer Automated Expert Semi-Analytical Resummation) 
to establish these properties with the
desired degree of reliability and precision, we have found it useful
to make use of multiple-precision arithmetic\cite{MP}.

\subsection{Master formula}
Given the above conditions, one can derive the following NLL master
resummation formula for the distribution $\Sigma(v)$\cite{autolet+autopaper}
\begin{eqnarray}
  \label{eq:Master}
 & &\!\!\!\!\ln \Sigma(v)=\!-\sum_{\ell=1}^n C_\ell \left[r_\ell(L) + 
    r_\ell'(L) \left(\ln {\bar d}_\ell - b_\ell \ln
      \frac{2E_\ell}{Q}\right)
    +   B_\ell \, T\!\left(\frac{L}{a+b_\ell}\right)
  \right] \nonumber \\
  &&+\sum_{\ell=1}^{n_i} \ln \frac{f_\ell(x_\ell,v^{\frac{2}{a+b_\ell}}
    \muf^2)}{f_\ell(x_\ell, \muf^2)} 
  + \ln S\left(T(L/a\right)) 
  + \ln \cF(C_1 r_1',\ldots,C_n r_n'), 
\end{eqnarray}
where $C_\ell$ is the color factor associated with Born leg $\ell$,
and $E_\ell$ is its energy, $B_\ell$ is $-3/4$ for quarks and $-(11\CA
- 4\TR \nf)/(12\CA)$ for gluons, $\ln {\bar d}_\ell = \ln d_\ell +
\int_0^{2\pi} \frac{d\phi}{2\pi} \ln g_\ell(\phi)$, and for incoming
legs, $f_\ell$ are the parton densities.
We note that \eqref{eq:Master} is independent of the frame in which
one determines the $d_\ell$ and (to NLL accuracy) of the choice of
hard scale $Q$.

\noindent 
The functions $r_\ell(L)$, which contain all the LL (and some NLL) terms, are
\begin{multline}
  \label{eq:rad-dl}
  r_\ell(L) \!=\!  \int_{v^{\frac2a}Q^2}^{ v^{\frac{2}{a+b_\ell}}Q^2
  }\!\frac{dk_t^2}{k_t^2}\frac{\as(k_t)}{\pi}
  \ln\left(\frac{k_t}{v^{1/a}Q}\right)^{\frac{a}{b_\ell}}\!  + \int_{
    v^{\frac{2}{a+b_\ell}}Q^2 }^{Q^2}
  \frac{dk_t^2}{k_t^2}\frac{\as(k_t)}{\pi}\ln\frac{Q}{k_t}\>,
\end{multline}
here $\as$, in the Bremsstrahlung scheme\cite{CMW}, runs at two-loop
order. 
$r_\ell' = \partial_L r_\ell$ and $T(L)$ are  relevant only at NLL
level
\begin{equation}
  \label{eq:T}
  T(L) = \int_{e^{-2L}Q^2
}^{Q^2}\frac{dk_t^2}{k_t^2}\frac{\as(k_t)}{\pi}\>.
\end{equation}

The process dependence associated with large-angle soft radiation is
described by $S(T(L/a))$, whose form depends on the number $n$ of
legs.  For $n=3$
\begin{equation}
\label{eq:lnSn=3}
\ln S(t) = -t \left[{\CA}\ln \frac{Q_{qg} Q_{q'
        g}}{Q_{q q'} Q}  
+ 2\CF \ln \frac{Q_{q q'}}{Q}\right] \,,
\end{equation}
where $Q^2_{ab} = 2p_a . p_b$ and $q$, $q'$ and $g$ denote the
(anti)-quarks and gluon. The simpler case $n=2$ can be read from
\eqref{eq:lnSn=3} setting $C_A =0$.  The $n=2,3$ formulae apply then
to $\ee$, DIS and Drell-Yan production, while a process such as $gg\to
\mathrm{Higgs} +g$ involves simply different color factors.
The (more involved) $n=4$ case needed to describe hadronic dijet
production can be found in\cite{autolet+autopaper}.

We examine now the factor $\cF$. Without it, eq.~\eqref{eq:Master}
corresponds essentially to the probability of vetoing all emissions
$k$ with $V(\{\tilde p\},k) > v$.  However at NLL subtle effects
enter: a simple veto on {\em single emissions} turns out to be
insufficient, since events might have $V(\{\tilde p\},k_1,\ldots,k_m)
> v$ though all emissions separately had $V(\{\tilde p\},k_{i}) < v$,
or vice versa.  This {\em multiple emission} effect, encoded in $\cF$,
is connected to how all secondary emission coherently determine the
value of the observable and can be computed in a general way
in\cite{BSZ}.

While this talk concentrates on the method, new results obtained with
it were presented at this conference in\cite{BanfiTalk} (and other
results can be found in\cite{autolet+autopaper}).

\section{Final remarks}
This project started with the development of an algorithm to
numerically compute the non-trivial NLL terms associated with multiple
emissions. It was initially applied by hand to three new observables
[6]. Progress since then includes the derivation of a general master
formula, together with a precise, automatically verifiable list of
conditions on the observable for the resummation to be valid at NLL
and the full numerical implementation of this.
Though no human intervention is needed, results have the quality of
analytic NLL predictions, so that any hadronisation model can be
applied, studies of renormalization and factorization scale dependence
can be carried out and a matching with fixed order is feasible.
The most import results obtained up to now are the first resummations
in hadronic dijet production, for a variety of observable
at a time.  
We now aim at automating the matching with fixed order
results\cite{NLOJET}, this will open up the possibility to carry out a
vast amount of phenomenological studies.

\section*{Acknowledgements} 
I thank the organizers of DIS2003, in particular Yuri Dokshitzer, for
the friendly atmosphere at the conference, and for the opportunity to
visit a wonderful town.

\end{document}